# Investigation of the role of polysaccharide in the dolomite growth at low temperature by using atomistic simulations

Zhizhang Shen[1], Philip E. Brown[1], Izabela Szlufarska*[2,3] and Huifang Xu*[1,3]

[1] NASA Astrobiology Institute, Department of Geoscience, University of Wisconsin-Madison, 1215 W. Dayton Street, Madison, WI53706, USA

[2] Department of Materials Science and Engineering, University of Wisconsin-Madison, 1509 University Avenue, Madison, WI53706, USA

[3] Materials Science Program, University of Wisconsin-Madison, 1509 University Avenue, Madison, WI53706, USA

*Corresponding authors

szlufarska@wisc.edu

Tel: 608-265-5878  Fax: 608-262-8353

hfxu@geology.wisc.edu

Tel: 608-265-5887  Fax: 608-262-0693

ABSTRACT

Dehydration of water from surface $Mg^{2+}$ is most likely the rate-limiting step in the dolomite growth at low temperature. Here, we investigate the role of polysaccharide in this step using classical molecular dynamics (MD) calculations. Free energy (potential of mean force, PMF) calculations have been performed for water molecules leaving the first two hydration layers above the dolomite (104) surface under the following three conditions: without catalyst, with monosaccharide (mannose) and with oligosaccharide (three units of mannose). MD simulations reveal that there is no obvious effect of monosaccharide in lowering the dehydration barrier for surface $Mg^{2+}$. However, we found that there are metastable configurations of oligosaccharide, which can decrease the dehydration barrier of surface $Mg^{2+}$ by about 0.7-1.1 kcal/mol. In these configurations, the molecule lies relatively flat on the surface and forms a bridge shape. The hydrophobic space near the surface created by the non-polar –CH groups of the oligosaccharide in the bridge conformation is the reason for the observed reduction of dehydration barrier.

**Introduction**

Dolomite ($CaMg(CO_3)_2$) is a very common sedimentary carbonate mineral that has a similar structure to calcite ($CaCO_3$), but with an alternating Ca and Mg basal layers. The uneven temporal distribution of dolomite abundance through the geological records has been bewildering geologists for many decades[1]. To solve this so called "dolomite problem", it is essential to understand the mechanisms of dolomite formation. In addition to studies of dolomitization mechanisms[1–4], extensive number of papers have been reported on synthesizing dolomite in ambient laboratory environments[5–9]. As a result, it is now recognized by the community that the growth of dolomite at low temperature (<40°C) cannot be accomplished without catalysts[10]. In general, these catalysts can be divided into two groups: 1) anaerobic bacteria that exist in modern environments where dolomite is precipitating[1,5–9]; 2) metabolites, exudates or cellular components that are related to those anaerobic bacteria, among which dissolved hydrogen sulfide and polysaccharides have shown to be very effective catalysts[11–13]. Despite the progresses made in searching for eligible catalysts, the role of the catalysts in the dolomite formation remains elusive. For instance, it is unknown whether the catalysts take effect in the solution or on the dolomite surface.

Before investigating the role of the catalysts in detail, it is important to first identify the rate-limiting step during the dolomite formation processes. The prevalent opinion[14] is that the key factor accounting for the sluggish kinetics of dolomite growth at room temperature is the Mg hydration. Mg hydration may inhibit the dolomite formation in two ways that are illustrated in Figure 1: 1) it is difficult for hydrated $Mg^{2+}$ ions to be adsorbed onto the dolomite surface from solution (i.e., from step 1 to step 2 in Figure 1); 2) the hydration layers above the surface block the binding of carbonate to surface $Mg^{2+}$ ions (from step 3 directly to step 7 in Figure 1). Previous simulations have shown that the adsorption of hydrated $Mg^{2+}$ ions is not a rate-limiting step and that hydrogen sulfides have a negligible effect on the dehydration of $Mg^{2+}$-$H_2O$ complex in solution [15–17]. As for the adsorption of carbonate to the surface $Mg^{2+}$ ions, two mechanisms have been proposed to enable this process: one is to remove the surface water bound to surface $Mg^{2+}$ or to weaken the surface $Mg^{2+}$-$H_2O$ interactions so that carbonate ions are able to approach surface $Mg^{2+}$ (the pathway from step 3 to 4 to 5 and finally to 7 as shown in Figure 1); the second mechanism[14] is to increase the availability of carbonate ions (i.e., alkalinity) so that carbonate can compete with surface water for binding to surface $Mg^{2+}$ (pathway from step 3 to 6 and then to 7 in Figure 1). However, the latter mechanism is not supported by experiments[11,12]. Thus, it is most likely that catalysts play an active role in dehydration of surface $Mg^{2+}$, but the question is how. One hypothesis for the dehydration of surface $Mg^{2+}$ is that catalysts can be strongly adsorbed to crystal surface replacing the surface water molecules[11,12]. Our previous study of the hydrogen sulfide adsorption on dolomite surface does not support this idea and suggests that water is more favorably adsorbed onto the surface than the hydrogen sulfide is[18]. Another hypothesis is that catalysts can lower the dehydration barrier of surface $Mg^{2+}$ so that carbonate can approach and be adsorbed onto the surface. The goal of this study is to test the latter hypothesis.

Polysaccharides were chosen for our study because, as mentioned above, this class of molecules has been shown to promote the dolomite growth at room temperature [9]. It is also because polysaccharides are the major component of the extracellular polymeric substance (EPS) and EPS is a universal catalyst no matter what the bacteria species is. Instead of studying the long-chained polysaccharides that contain thousands of atoms, we tested α-D-mannose (one of the major monomers making up the polysaccharides used in experiments[12,13]) and oligo-mannose. Therefore, our simulations have only considered the local small blocks of the long chained polysaccharides. The effects of mannose and oligo-mannose in the dehydration barrier of surface $Mg^{2+}$ were investigated by using classical MD simulations and PMF method.

**Method**

**Force fields for molecular dynamics simulations**

Force fields for $CaCO_3$, which include a polarizable core-shell model to account for the electronic polarizability of oxygen atoms in $CO_3^{2-}$ group, were developed by Pavese et al. (1996)[19]. Force fields for $MgCO_3$ were developed by de Leeuw and Parker, 2000[20]. However, a recent work has shown that a rigid-ion model of calcite can provide a better description of the interfacial (calcite-water) water structure than the polarizable model[21]. In this rigid-ion version of the calcite force field, the shells are removed and the oxygen charge is the sum of the core and shell while all other parameters remain the same as in the Pavese model[19]. We followed the same method to construct a rigid-ion force field for dolomite. Details of the force field parameters can be found in the Appendix.

For mannose (and tri-mannose) and water we used glycam_06h[22] and TIP3P[23] force fields, respectively. The parameters of mannose and tri-mannose, water and the interactions between them were obtained from DL_FIELD_3.1[24]. The initial mannose and tri-mannose structures were prepared using Carbohydrate Builder code[25]. For dolomite-water interactions, we used the following parameters: the calcium-water oxygen/magnesium-water oxygen interactions were taken from de Leeuw and Parker (2000)[20]; the carbon-water oxygen interactions were adapted from Kerisit et al. (2003)[26].

Because of the large charge disparity between dolomite and sugar, it is important to derive the cross-term force fields so that they are consistent with the Coulomb interactions between dolomite and sugar atoms[27]. By following the method described by Freeman et al. (2007)[27], we fitted the $A$ parameters in Buckingham potentials for Ca-$O_{sugar}$ and Mg-$O_{sugar}$ interactions together to the dolomite lattice parameters. The Buckingham potential can be expressed as: $U(r) = Aexp\left(-\frac{r}{\rho}\right) - \frac{c}{r^6}$, where $A$ and $\rho$ are related to the number of electrons and the electron density, respectively. The first term in $U(r)$ describes the short-range repulsive interactions, while the second term describes attraction by van der Waals forces. The optimal combination of 1000 eV and 650 eV was obtained for $A$ parameters in Ca-$O_{sugar}$ and Mg-$O_{sugar}$ interactions, respectively. The Coulomb interactions between other non-bonded atoms are usually very weak compared to the interactions between $Ca^{2+}$/$Mg^{2+}$ and $O_{sugar}$ and for these cross-terms we used parameters derived in previous studies[27]. In order to validate the fitting parameters, the most stable configuration of mannose on dolomite (104) surface identified from MD simulations at 10K was then relaxed in *ab initio* calculations using the density functional theory (DFT) as implemented in VASP[28]. Details of VASP calculations can be found in the Appendix. The center of mass of mannose in classical MD simulations was found to be closer to the surface by 0.3Å as compared to the DFT results, which provides a reasonable agreement. The bond lengths of Ca-O and Mg-O in MD simulations were found to be 2.41Å and 2.04Å, respectively, while the corresponding values from VASP are 2.48 Å and 2.28 Å. Although there are some quantitative differences between the values obtained with classical force fields and DFT, these differences are relatively small and the qualitative trends in bonding are in agreement between the two approaches. In addition, the overall structure and orientation of mannose are similar in both MD and DFT simulations.

**Simulation Details**

The dolomite slab has approximate dimensions of 23.12×24.06×19.10 Å$^3$. The slab has two (top and bottom) free surfaces with the (104) orientation. The slab contains 105 calcium, 105 magnesium, 210 carbon and 630 oxygen atoms and it has seven atomic layers with atoms in the middle three layers of the slab being fixed in space while the top two and the bottom two layers are allowed to relax during simulations. A water box is added above the top surface and because periodic boundary conditions are applied in all three spatial dimensions, this water box is also in contact with the dolomite bottom surface. The total height of the water box is 52 Å. The water box has 880 water molecules, which corresponds to the density of ~ 0.9 g/cm$^3$. In the simulations of water adsorption, mannose and tri-mannose adsorptions were first relaxed on dolomite (104) surface by using classical MD simulations at room temperature and 1 atmosphere. The code DL_POLY_class[29] was used and each system was simulated in the NVE (constant number of particles, volumes, and energy) ensemble. Equations of motion were integrated using the Verlet-Leapfrog algorithm using a timestep of 1 fs. The long-range Coulomb interactions were calculated using the smooth particle mesh Ewald sum (SPME) with a real space cut-off of 10.1 Å. Multiple configurations of mannose and tri-mannose were explored and the details regarding the configurations can be found in the results section.

The potential of mean force (PMF) calculations were carried out for water molecules binding/unbinding to Mg on dolomite (104) surface, both in the presence and in the absence of sugar. The reaction coordinate is the distance between the oxygen atom of a water molecule and a surface Mg atom along the z direction. The PMF calculations were conducted in NVT (constant number of particles, volumes, and temperature) ensemble by using the Nosé-Hoover thermostat with a relaxation time of 0.5 ps. The PLUMED package[30] interfaced to DL_POLY Classic was employed to run Umbrella Sampling[31,32]. The weighted histogram analysis method (WHAM[33]) was used to analyze the results of umbrella sampling and to calculate the unbiased free energy along the reaction pathway. In each sampling window, we ran simulations for 6 ns, which was sufficient to reach the convergence of unbiased free energy within 0.1 kcal/mol. Multiple simulations were conducted in each case, as explained in the subsequent sections.

**Results**

In order to determine the effects of sugar on the dehydration of surface Mg, we first studied water adsorption on dolomite (104) surface in the absence of sugar and we used it as a reference for later calculations that involve sugar. Classical MD simulation of the system was performed for 2ns at room temperature and 1 atmosphere to examine the water configurations on the dolomite surface. Radial distribution functions (RDF) of Ca-$O_{water}$, Mg-$O_{water}$ and $O_{carbonate}$-$H_{water}$ found peaks at 2.38 Å, 1.9 Å and 1.65 Å, respectively (Figure 2). A snapshot of the MD simulation of the water-dolomite (104) surface is shown in Figure 3a. Based on our analysis of water density profile along the z direction (Figure 3b), we can see that there are two hydration layers above the dolomite surface: one is at ~2.0 Å above the average height of the surface cations, and another one at 3.06 Å. The first and second peaks at 1.88 Å and 2.23 Å in the density profile correspond to the interactions between surface Mg and O atoms in water molecules and between Ca and water O atoms, respectively. The peak at 3.06 Å represents interactions between water hydrogen and oxygen of surface. An x-ray reflectivity work of water-dolomite (104) interface found a first hydration layer at a height of 2.31 Å above the surface cation[34]. The two-layer hydration is similar to that found at water-calcite (104) interface, which has been studied more extensively than water/dolomite surfaces. Discrepancies in the height of the two-hydration layers between simulations and experiments are not uncommon in calcite-water interface studies. For example, the height of the first hydration layer was determined from experiments to be at 2.0 Å[35] or 2.14 Å[36] while the simulations show a range from 2.0 to 2.4 Å[26,37,38]. The discrepancies found for dolomite/water interface between our simulation and the scarce previous experimental studies are of the similar order of magnitude as for the calcite/water interface. In our simulations we identified two types of metastable water configurations above surface Mg. In the first configuration one of the hydrogen atoms of water is pointing towards the nearby carbonate due to a hydrogen bond between the hydrogen atom and the oxygen atom in the carbonate ion (Figure 4a). In the second configuration (Figure 4b) both hydrogen atoms point upwards (away from the surface) and they are not bonded to the surface carbonate. Four water molecules (two with configuration 1 and two with configuration 2) were chosen for the subsequent calculations of free energies. Specifically, using methods described in the previous sections we calculated PMF of a water molecule being desorbed from the surface through the first two hydration layers into the bulk. In Figures 4c and 4d, we can see that there are two energy wells at ~1.9 Å and ~3.2 Å, which correspond to the two hydration layers, respectively. Free energy of water molecules residing in the first hydration layer varies from -6.67 to -7.32 kcal/mol. Here, we find that the energy barrier for dissociation of a water molecule from the surface Mg atoms (equivalent to the energy barrier to move water molecule away from the surface Mg past the first hydration layer) ranges from 9.40 ± 0.12 to 9.69 ± 0.10 kcal/mol. This range is close to energies calculated for a water molecule leaving the first hydration shell of $Mg^{2+}$ in solution, which are 9.0 kcal/mol[16] or ~10.5 kcal/mol[17].

To examine whether polysaccharides could decrease the hydration barrier of surface water, we first tested the effect of monosaccharide (here mannose), which is the basic unit of carbohydrate and which contains both the hydrophilic –OH group and the non-polar C–H bonds. Four different mannose configurations on the dolomite surface were simulated and two of the relaxed structures were chosen for PMF calculations based on the fact that they have the lowest energy configurations among the four initial configurations. The two lowest energy configurations represent two distinct orientations of mannose on the surface: one lies flat on the surface (Figures 5a and 5b) and the other one stands vertically above the surface (Figure 5c). In the system with mannose lying flat on the surface, two sets of PMF calculations were performed for the two water molecules that are adjacent to the mannose and that are bonded to the surface Mg (Figures 5a and 5b). The energy barriers for the first layer hydration are found to be 9.83 ± 0.10 and 9.53 ± 0.16 kcal/mol (Figures 5d and 5e), which means that there is no obvious effect of mannose with flat configuration on the dehydration of surface Mg. Interestingly, in one of the PMF profiles (Figure 5d), the peak at ~3.7 Å (~10.5 kcal/mol) is much higher than the remaining PMF profile, indicating an increase of dehydration barrier for the second hydration layer. This increase is caused by the interactions between hydroxyl group of the mannose and the desorbing water molecule as the molecule is trying to leave the second hydration layer. Qualitatively similar results are found in the system with a vertical mannose, where the energy barrier for the water molecule leaving the surface Mg (9.67 ± 0.13 kcal/mol) (Figure 5f) is not significantly different from the case without mannose.

Although the monomer of mannose was found to have a negligible effect on the dehydration of surface Mg-$H_2O$ complex, it is possible that the polymerization of monomers plays an important role in the process. To test this hypothesis, we investigate oligo-saccharide (or tri-saccharide, three units of α-D-monosaccharide joined by 1-4 linkage) as a model polysaccharide. Seven different tri-mannose configurations above dolomite (104) surface were explored by performing MD simulations for 3 ns at room temperature and 1 atmosphere (see appendix for the details and energies of the seven configurations). In the lowest energy system (configuration 1), one unit (ring) of the three six-member rings is bound to the surface while the other two units are pointing away from the surface. The configuration of the unit that bonds to the surface and the solvation environment around this unit are very similar to the system of a vertical mannose adsorbed onto the surface. Consequently, we do not expect this configuration of tri-mannose to have a significant weakening effect on the surface Mg-water interactions. The energies of the other six configurations of tri-mannose systems are comparable to each other. In configuration 2, two units of the tri-mannose bond to the dolomite surface and the third unit points away from the surface. The other metastable configurations (3,4, 5, 6 and 7), each with energy about 1~2 eV higher for the entire simulation box (containing more than 3000 atoms) than the ground state configuration 1, display a bridge shape but overall lie flat on the surface (Figures 6a and 6b). These bridge-shaped configurations create low water density space just beneath the bridge near the hydrophobic –CH groups. The effect of this bridge shape and the space of low water density under the bridge cannot be tested in the case of monomers. Thus, we performed PMF calculations of desorption of two surface water molecules for configurations 3 and 5 (four PMF calculations in total). In one case, the water molecule lies beneath the bridge and the low water density space is above Mg ions. In another case the desorbing water molecule lies outside the bridge but is adjacent to the low water density space. In the PMF profiles (Figure 6c) for water molecules under the bridge, we find a decreased energy barrier for the dehydration of the first hydration layer. This barrier is ~8.67 ± 0.12 kcal/mol in configuration 3 and ~8.59 ± 0.11 kcal/mol in configuration 5, which is about 0.73 ± 0.17 ~ 1.1 ± 0.14 kcal/mol lower than the corresponding barrier in pure water adsorption system. The PMF profile for the water molecule just outside the bridge (Figure 6d) has shown a similar amount of decrease in the hydration barrier (~8.66 ± 0.14 kcal/mol).

**Discussion**

Our PMF calculations show that the monosaccharide (mannose) does not have the ability to promote the dehydration of surface magnesium. Although the calculations were carried out with the presence of one monosaccharide, the concentration of monosaccharide is not expected to be a major factor. According to our complementary experiments at low temperature (see appendix), the increase of monosaccharide used in the synthesis does not foster the incorporation of Mg into the carbonate structure.

We considered the two major metastable configurations of tri-mannose on the dolomite (104) surface (one standing up vertically and one lying flat). The bridge-shaped configurations of oligosaccharide lying flat on the dolomite surface are shown to be able to decrease the dehydration barrier for surface magnesium. Hydrophobic space with low water density created by this bridge shape is also observed in our MD simulations of solvation of other tri-saccharides (tri-glucose, tri-galactose and tri-xylose). This phenomenon is common in the helix of polysaccharides where the interior of the helix is hydrophobic while the outer shell is hydrophilic[39]. The bridge-shaped configurations are not the lowest energy configurations that we explored in our study but the energy difference is small considering the number of atoms in the simulation box. The flat configuration could be further stabilized considering the following scenario. In EPS, the polysaccharides usually contain substituents of low molecular weight, such as negatively charged carboxylate groups and sulfates[40]. In aqueous environment, the carboxylate groups can replace the water molecules bonded to $Mg^{2+}$ and $Ca^{2+}$ ions[41]. For example, the Log K values (the logarithms of the equilibrium quotients) for the following two reactions at zero ionic strength are both 1.43:

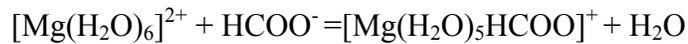
$$[Mg(H_2O)_6]^{2+} + HCOO^- = [Mg(H_2O)_5HCOO]^+ + H_2O$$

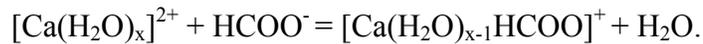
$$[Ca(H_2O)_x]^{2+} + HCOO^- = [Ca(H_2O)_{x-1}HCOO]^+ + H_2O.$$

Because of the strong attractions between Ca/Mg and carboxylate/sulfate, the flat configuration of the EPS polysaccharides may be further stabilized and perhaps even become the ground state structure on the dolomite surface.

Most of the previous studies on the role of polysaccharides or carboxyl groups in dolomite growth have been focused on the desolvation of $[Mg(H_2O)_6]^{2+}$ in the solution and the adsorption of Mg to the surface. One mechanism explored by some recent works is one where the surface carboxyl functional groups dewater and adsorb $Mg^{2+}$ and subsequently form a thin dolomite template for further precipitation under supersaturated conditions[9,42]. Although the energy barrier for the dehydration of one water molecule from the $[Mg(H_2O)_6]^{2+}$ is about 9.0~10.5 kcal/mol[16,17], the energy barrier for the adsorption of hydrated $Mg^{2+}$ ions on the calcite surface is only 3.06 kcal/mol and is very similar to the adsorption of calcium[16]. Therefore, the adsorption of hydrated Mg from solution to the surface is not likely a rate-limiting step for the dolomite growth. In addition, a thin dolomite template does not necessarily facilitate further growth of dolomite. For instance, X-ray reflectivity and atomic force microscopy experiments have shown that only 1-layer[43] or 2-layer[30] thick film could be formed on the dolomite substrate during reaction with the supersaturated solution. Furthermore, a careful examination of the results from Refs[9,42,44] indicates that the published XRD data are too noisy to uniquely identify dolomite peaks[45].

Our simulations have focused on the dehydration of surface Mg and we have shown that the polysaccharide is able to decrease the dehydration barrier for the surface Mg by its hydrophobic –CH groups, which is a possible reason why polysaccharide is important in the dolomite growth at low temperature. Nevertheless, the pathway for carbonate ions to supplant the surface water molecules in the presence of polysaccharide is complex and needs to be further investigated.

**REFERENCES**

(1) Warren, J. Dolomite: Occurrence, Evolution and Economically Important Associations. *Earth-Science Rev.* **2000**, *52*, 1–81.

(2) Adams, J. E.; Rhodes, M. L. Dolomitization by Seepage Refluxion. *Am. Assoc. Pet. Geol. Bull.* **1960**, *44*, 1912–1920.

(3) Badiozamnai, K. The Dorag Dolomitization Model - Application to the Middle Ordovician of Wisconsin. *J. Sediment. Res.* **1973**, *43*, 965–984.

(4) Hardie, L. A. Dolomitization; a Critical View of Some Current Views. *J. Sediment. Res.* **1987**, *57*, 166–183.

(5) Vasconcelos, C.; McKenzie, J. Microbial Mediation of Modern Dolomite Precipitation and Diagenesis under Anoxic Conditions (Lagoa Vermelha, Rio de Janeiro, Brazil). *J. Sediment. Res.* **1997**, *67*, 378–390.

(6) Van Lith, Y.; Warthmann, R.; Vasconcelos, C.; Mckenzie, J. a. Sulphate-Reducing Bacteria Induce Low-Temperature Ca-Dolomite and High Mg-Calcite Formation. *Geobiology* **2003**, *1*, 71–79.

(7) Roberts, J. A.; Bennett, P. C.; González, L. A.; Macpherson, G. L.; Milliken, K. L. Microbial Precipitation of Dolomite in Methanogenic Groundwater. *Geology* **2004**, *32*, 277–280.

(8) Wright, D. T.; Wacey, D. Precipitation of Dolomite Using Sulphate-Reducing Bacteria from the Coorong Region, South Australia: Significance and Implications. *Sedimentology* **2005**, *52*, 987–1008.

(9) Kenward, P. A.; Goldstein, R. H.; González, L. A.; Roberts, J. A. Precipitation of Low-Temperature Dolomite from an Anaerobic Microbial Consortium: The Role of Methanogenic Archaea. *Geobiology* **2009**, *7*, 556–565.

(10) Land, L. S. Failure to Precipitate Dolomite at 25° C fromDilute Solution Despite 1000-Fold Oversaturation after32 Years. *Aquat. Geochemistry* **1998**, *4*, 361–368.

(11) Zhang, F.; Xu, H.; Konishi, H.; Kemp, J. M.; Roden, E. E.; Shen, Z. Dissolved Sulfide-Catalyzed Precipitation of Disordered Dolomite: Implications for the Formation Mechanism of Sedimentary Dolomite. *Geochim. Cosmochim. Acta* **2012**, *97*, 148–165.

(12) Zhang, F.; Xu, H.; Konishi, H.; Shelobolina, E. S.; Roden, E. E. Polysaccharide-Catalyzed Nucleation and Growth of Disordered Dolomite: A Potential Precursor of Sedimentary Dolomite. *Am. Mineral.* **2012**, *97*, 556–567.

(13) Zhang, F.; Xu, H.; Shelobolina, E. S.; Konishi, H.; Converse, B.; Shen, Z.; Roden, E. E. The Catalytic Effect of Bound Extracellular Polymeric Substances Excreted by Anaerobic Microorganisms on Ca-Mg Carbonate Precipitation : Implications for the " Dolomite Problem ." *Am. Mineral.* **2015**, *100*, 483–494.

(14) Lippmann, F.; Lippmann, F. *Sedimentary Carbonate Minerals*; Springer, 1973; Vol. 228.

(15) De Leeuw, N. H.; Parker, S. C. Surface–water Interactions in the Dolomite Problem. *Phys. Chem. Chem. Phys.* **2001**, *3*, 3217–3221.

(16) Kerisit, S.; Parker, S. C. Free Energy of Adsorption of Water and Metal Ions on the [10 1 4] Calcite Surface. *J. Am. Chem. Soc.* **2004**, 10152–10161.

(17) Yang, Y.; Sahai, N.; Romanek, C. S.; Chakraborty, S. A Computational Study of Mg2+ Dehydration in Aqueous Solution in the Presence of HS− and Other

(18) Shen, Z.; Liu, Y.; Brown, P. E.; Szlufarska, I.; Xu, H. Modeling the Effect of Dissolved Hydrogen Sulfide on Mg 2+ – Water Complex on Dolomite {104} Surfaces. *J. Phys. Chem. c* **2014**, *118*, 15716–15722.

(19) Pavese, A.; Catti, M.; Parker, S. C.; Wall, A. Modelling of the Thermal Dependence of Structural and Elastic Properties of Calcite , CaCO 3. *Phys. Chem. Miner.* **1996**, *23*, 89–93.

(20) De Leeuw, N.; Parker, S. Modeling Absorption and Segregation of Magnesium and Cadmium Ions to Calcite Surfaces: Introducing MgCO3 and CdCO3 Potential Models. **2000**.

(21) Fenter, P.; Kerisit, S.; Raiteri, P.; Gale, J. D. Is the Calcite – Water Interface Understood ? Direct Comparisons of Molecular Dynamics Simulations with Specular X - Ray Reflectivity Data. *J. Phys. Chem. c* **2013**, *117*, 5028–5042.

(22) Kirschner, K. N.; Yongye, A. B.; Tschampel, S. M.; Gonza, J.; Daniels, C. R.; Foley, B. L.; Woods, R. J. GLYCAM06 : A Generalizable Biomolecular Force Field. Carbohydrates. *J. Comput. Chem.* **2007**, *29*, 622–655.

(23) Jorgensen, W. L.; Chandrasekhar, J.; Madura, J. D.; Impey, R. W.; Klein, M. L. Comparison of Simple Potential Functions for Simulating Liquid Water. *J. Chem. Phys.* **1983**, *79*, 926.

(24) Yong, C. W. DL _ FIELD – A Force Field and Model Development Tool for DL _ POLY. *CSE Front.* **2010**, 38–40.

(25) Woods Group. GLYCAM Web. Complex Carbohydrate Research Center, U. of G. Carbohydrate Builder Code.

(26) Kerisit, S.; Parker, S. C.; Harding, J. H. Atomistic Simulation of the Dissociative Adsorption of Water on Calcite Surfaces. *J. Phys. Chem. B* **2003**, *107*, 7676–7682.

(27) Freeman, C. L.; Harding, J. H.; Cooke, D. J.; Elliott, J. a.; Lardge, J. S.; Duffy, D. M. New Forcefields for Modeling Biomineralization Processes. *J. Phys. Chem. C* **2007**, *111*, 11943–11951.

(28) Kresse, G.; Furthmüller, J. Efficient Iterative Schemes for Ab Initio Total-Energy Calculations Using a Plane-Wave Basis Set. *Phys. Rev. B. Condens. Matter* **1996**, *54*, 11169–11186.

Monovalent Anions – Insights to Dolomite Formation. *Geochim. Cosmochim. Acta* **2012**, *88*, 77–87.

(29) Smith, W.; Forester, T. R. DL_POLY_2.0: A General-Purpose Parallel Molecular Dynamics Simulation Package. *J. Mol. Graph.* **1996**, *14*, 136–141.

(30) Bonomi, M.; Branduardi, D.; Bussi, G.; Camilloni, C.; Provasi, D.; Raiteri, P.; Donadio, D.; Marinelli, F.; Pietrucci, F.; Broglia, R. a.; Parinello, M. PLUMED: A Portable Plugin for Free-Energy Calculations with Molecular Dynamics. *Comput. Phys. Commun.* **2009**, *180*, 1961–1972.

(31) Kirkwood, J. G. Statistical Mechanics of Fluid Mixtures. *J. Chem. Phys.* **1935**, *3*, 300.

(32) Torrie, G. M.; Valleau, J. P. Monte Carlo Free Energy Estimates Using Non-Boltzmann Sampling: Application to the Sub-Critical Lennard-Jones Fluid. *Chem. Phys. Lett.* **1974**, *28*, 578–581.

(33) Kumar, S.; Bouzida, D.; Swedsen, R. H.; Kollman, P. A.; Rosenbergl, J. M. The Weighted Histogram Analysis Method for Free-Energy Calculations on Biomolecules. I. The Method. *J. Comput. Chem.* **1992**, *13*, 1011–1021.

(34) Fenter, P.; Zhang, Z.; Park, C.; Sturchio, N. C.; Hu, X. M.; Higgins, S. R. Structure and Reactivity of the Dolomite (104)-Water Interface: New Insights into the Dolomite Problem. *Geochim. Cosmochim. Acta* **2007**, *71*, 566–579.

(35) Fenter, P.; Lee, S. S. Hydration Layer Structure at Solid – Water Interfaces. *MRS Bull.* **2014**, *39*, 1056–1061.

(36) Fenter, P.; Sturchio, N. C. Calcite (104)-Water Interface Structure, Revisited. *Geochim. Cosmochim. Acta* **2012**, *97*, 58–69.

(37) Cooke, D. J.; Gray, R. J.; Sand, K. K.; Stipp, S. L. S.; Elliott, J. a. Interaction of Ethanol and Water with the {1014} Surface of Calcite. *Langmuir* **2010**, *26*, 14520–14529.

(38) Raiteri, P.; Gale, J. D.; Quigley, D.; Rodger, P. M. Derivation of an Accurate Force-Field for Simulating the Growth of Calcium Carbonate from Aqueous Solution: A New Model for the Calcite?water Interface. *J. Phys. Chem. C* **2010**, *114*, 5997–6010.

(39) Foster, J. F. Physical Properties of Amylose and Amylopectin in Solution. *Starch Chem. Technol.* **1965**, *1*, 349–392.

(40) Wingender, J.; Neu, T. R.; Flemming, H. C. What Are Bacterial Extracellular Polymeric Substances? In *Microbial Extracellular Polymeric Substances*; Springer Berlin Heidelberg, 1999; pp. 1–19.

(41) Martell, A. E.; Smith, R. M. *Critical Stability Constants*; Springer, 1974; Vol. 1.

(42) Roberts, J. A.; Kenward, P. A.; Fowle, D. A.; Goldstein, R. H.; González, L. A.; Moore, D. S. Surface Chemistry Allows for Abiotic Precipitation of Dolomite at Low Temperature. *Proc. Natl. Acad. Sci. U. S. A.* **2013**, 6–11.

(43) Higgins, S. R.; Hu, X. Self-Limiting Growth on Dolomite: Experimental Observations with in Situ Atomic Force Microscopy. *Geochim. Cosmochim. Acta* **2005**, *69*, 2085–2094.

(44) Kenward, P. a.; Fowle, D. a.; Goldstein, R. H.; Ueshima, M.; González, L. a.; Roberts, J. a. Ordered Low-Temperature Dolomite Mediated by Carboxyl-Group Density of Microbial Cell Watts. *Am. Assoc. Pet. Geol. Bull.* **2013**, *97*, 2113–2125.

(45) Gregg, J. M.; Bish, D. L.; Kaczmarek, S. E.; Machel, H. G. Mineralogy, Nucleation and Growth of Dolomite in the Laboratory and Sedimentary Environment: A Review. *Sedimentology* **2015**, in press.

Figure 1. Proposed steps in the dolomite growth at low temperature. Initially, $Mg^{2+}$ ions are in the solution and they are six-fold hydrated (step (1)). In step (2), the partially dehydrated $Mg^{2+}$ ions are adsorbed onto the dolomite surface. The retained water molecules make it difficult for carbonate ions to attach to the surface as shown in step (3). Two possible ways have been suggested to promote the adsorption of carbonate (from step (3) to step (7)). One way is to increase the availability of carbonate in the solution (step (6)), which hypothesis is not supported by experiments[11]. The second way is to add polysaccharide into the system, which can either replace the surface water or decrease the dehydration barrier so that it is easier for carbonate to approach to the surface $Mg^{2+}$, i.e. the step (4) and (5).

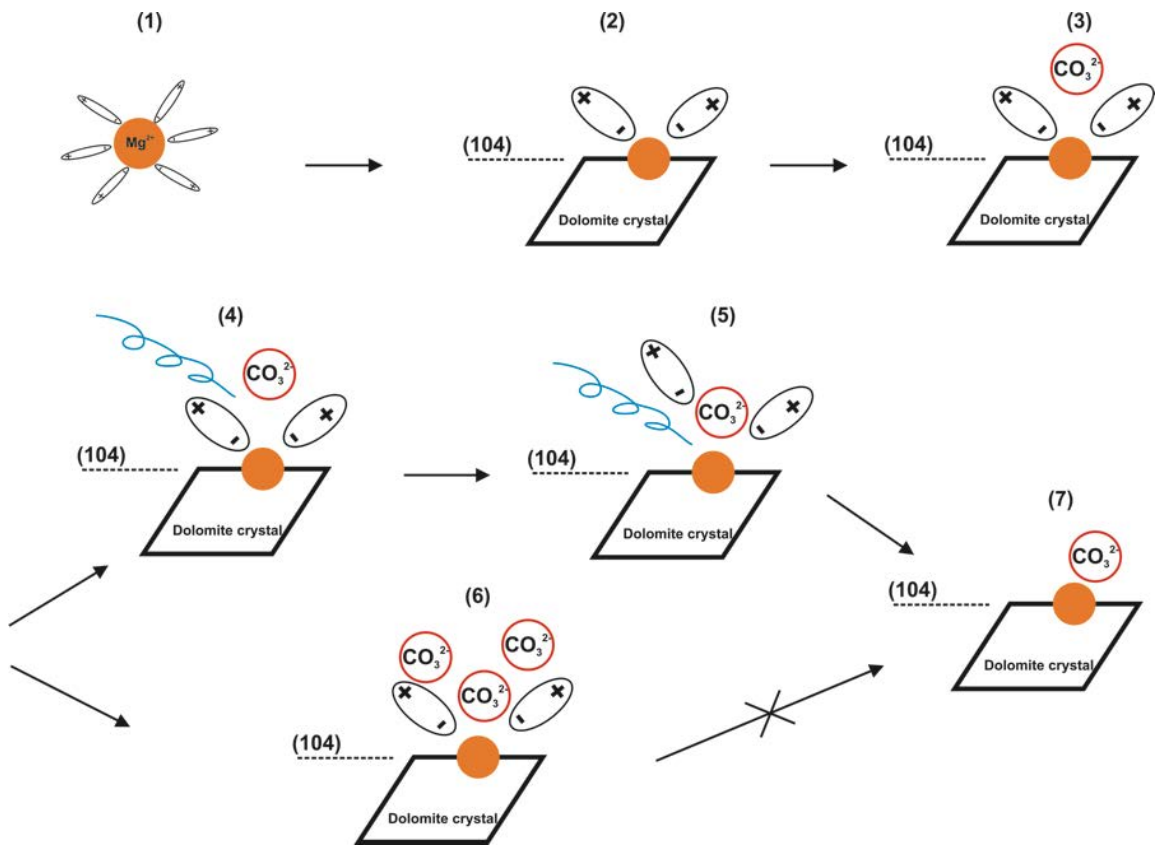

Figure 2. Radial distribution functions for Ca-Ow (water oxygen) pair (dark gray line), Mg-Ow pair (gray line) and H-O (carbonate oxygen) pair (black line).

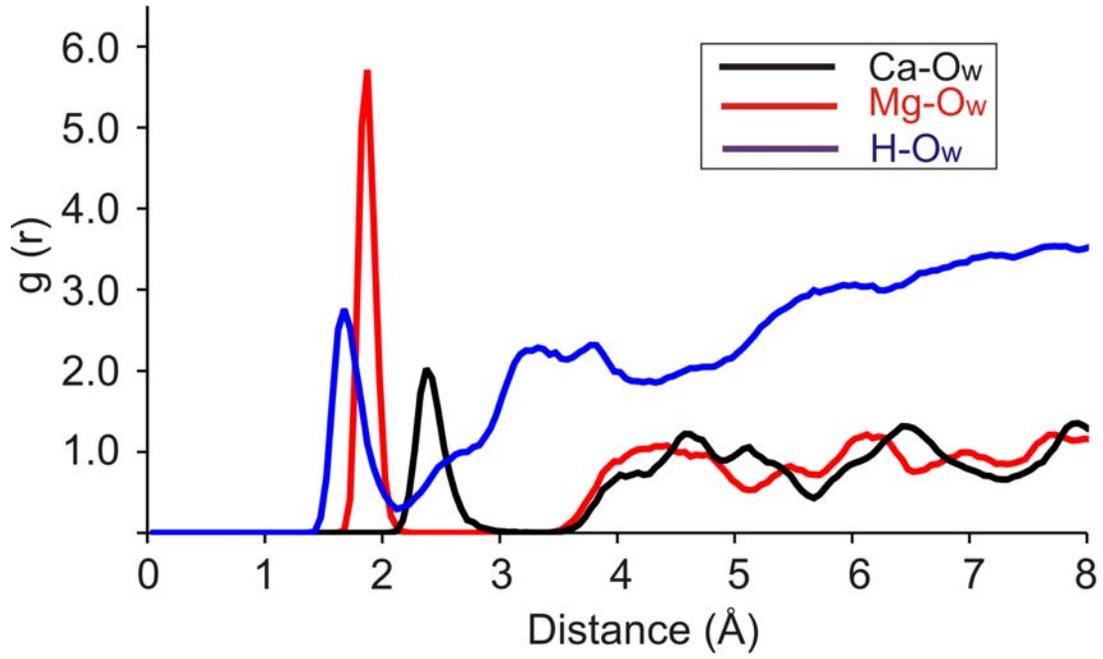

Figure 3. (a) Snapshot of MD simulation showing the interface between dolomite (104) surface and water molecules. The atoms Ca, Mg, C, O, Ow and H are in blue, range, brown, black and white colors. (b) The density profile of water molecules along the perpendicular distance from the surface.

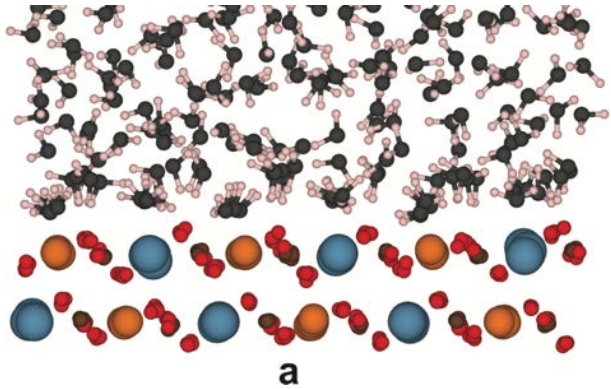

a

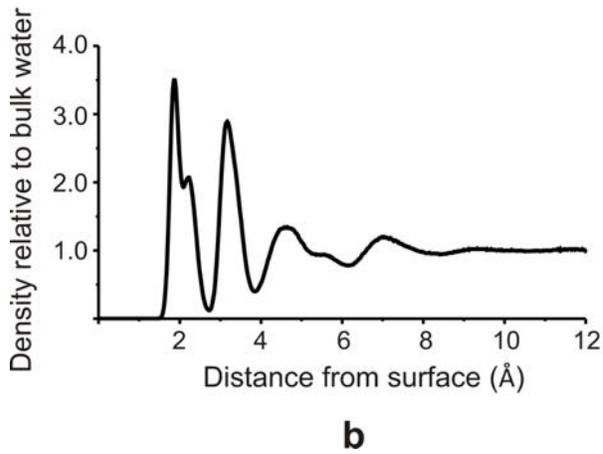

b

Figure 4. (a) The configuration of type 1 water molecules adsorbed onto surface $Mg^{2+}$. (b) The configuration of type 2 water molecules adsorbed onto surface $Mg^{2+}$. (c) Potential of mean force curve for water molecule I leaving the first two hydration layers. (d) Potential of mean force curve for water molecule II leaving the first two hydration layers.

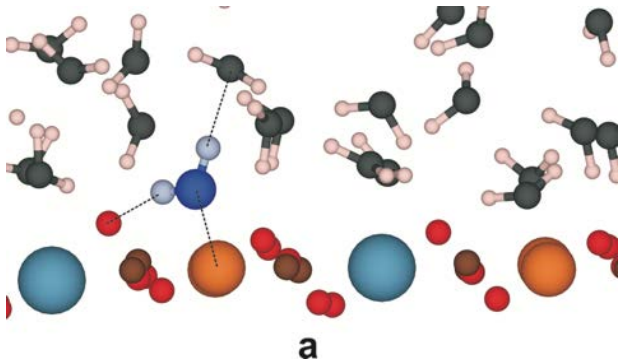

a

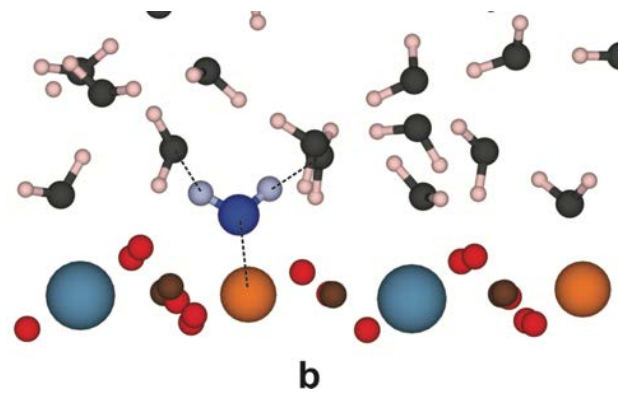

b

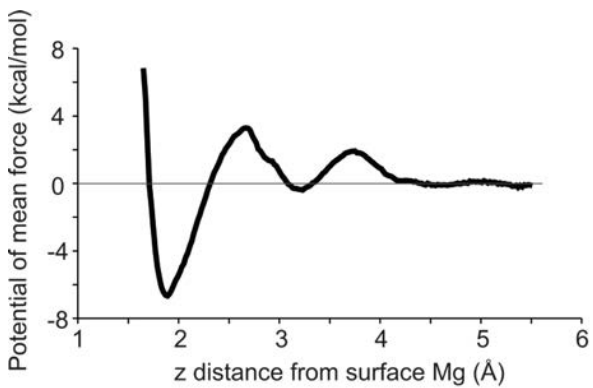

c

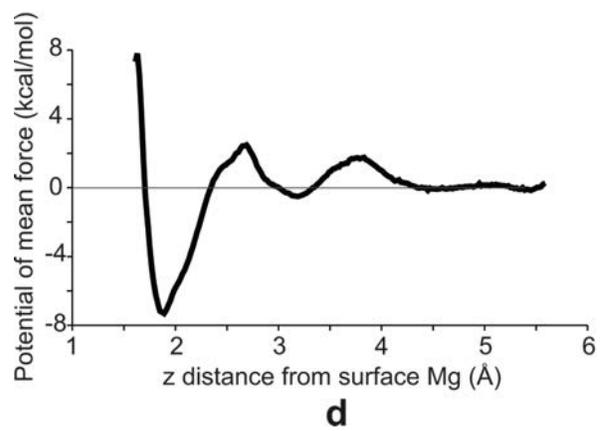

Figure 5. (a) A flat configuration of mannose adsorbed onto the surface. Water molecules are included in the simulations but not shown in this figure. (b) Side view of the mannose on the surface. (c) A vertical configuration of mannose adsorbed onto the surface. Water molecules are included in the simulations but not drawn in this figure. (d) Potential of mean force curve for water molecule #1 leaving the first two hydration layers in the presence of a flat mannose as shown in (a) and (b). (e) Potential of mean force curve for water molecule #2 leaving the first two hydration layers. (f) Potential of mean force curve for water molecule leaving the first two hydration layers in the presence of vertical mannose as shown in (c).

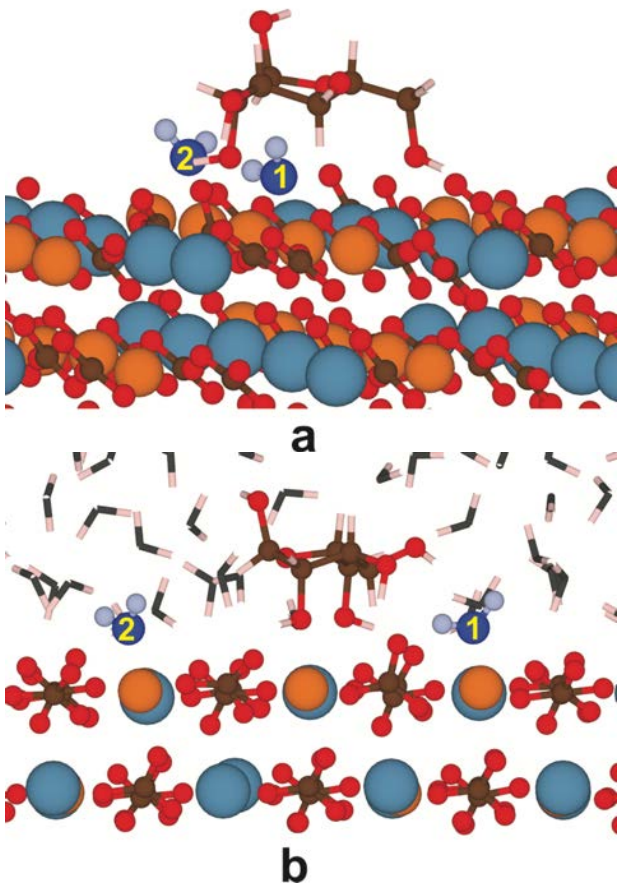

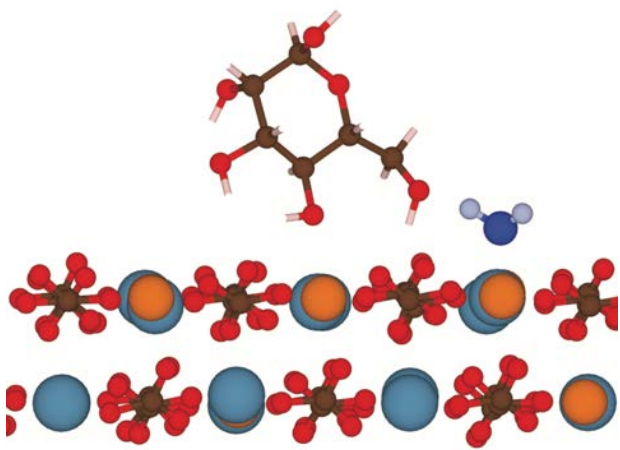

c

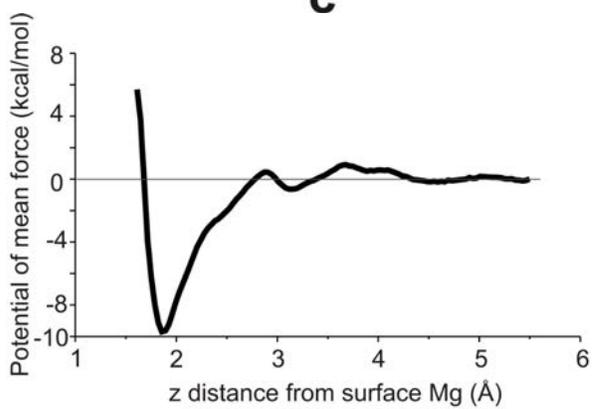

d

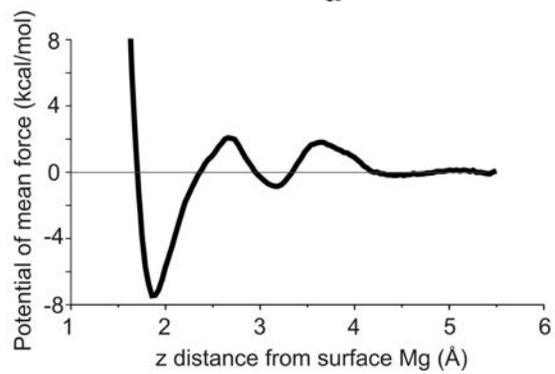

e

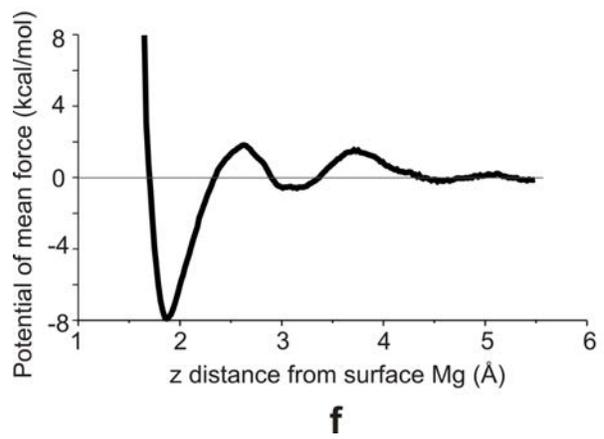

f

Figure 6. (a) A bridge shape configuration of tri-mannose lying flat on the dolomite surface. (b) Side view of the tri-mannose on the surface. Water molecules are included in the simulations but not shown in this figure. (c) Potential of mean force curve for water molecule #1 underneath the bridge leaving the first two hydration layers. (d) Potential of mean force curve for water molecule #2 outside the bridge leaving the first two hydration layers.

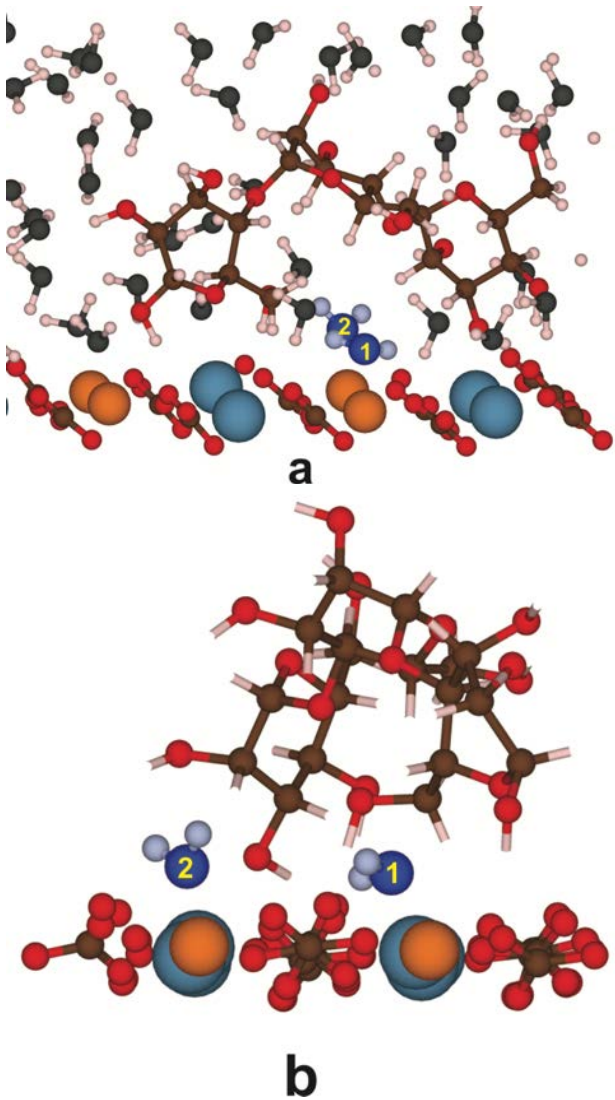

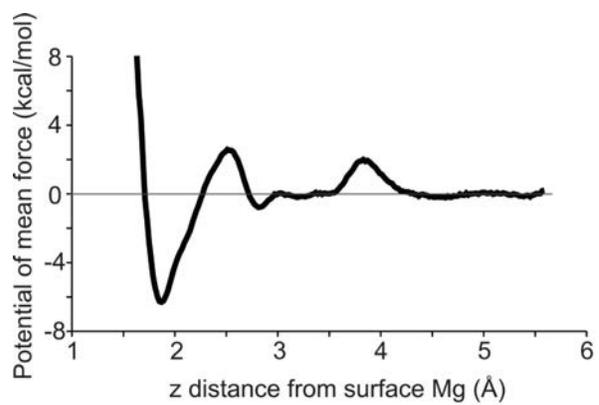

c

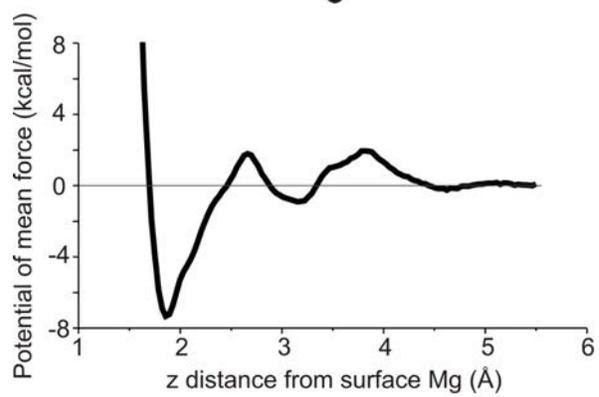

d